\documentclass[12pt,preprint]{aastex}
\usepackage{graphicx}
\def\ha{\relax \ifmmode {\rm H}\alpha\else H$\alpha$}

\def\halp11{\relax [KOS94]~HA\,11}
\def\ic{$I_\mathrm{C}$}
\def\rc{$R_\mathrm{C}$}
\def\ks{$K_\mathrm{s}$}

\begin{document}

\title{A peculiar young eruptive star in the dark cloud Lynds 1340}

\author{M. Kun\altaffilmark{1}, E. Szegedi-Elek\altaffilmark{1}, A. Mo\'or\altaffilmark{1}, 
\and P. \'Abrah\'am\altaffilmark{1}, J. A. Acosta-Pulido\altaffilmark{2}, D. Apai\altaffilmark{3}, 
J. Kelemen\altaffilmark{1}, A. P\'al\altaffilmark{1}, M. R\'acz\altaffilmark{1}, 
Zs. Reg\'aly\altaffilmark{1},  R. Szak\'ats\altaffilmark{4}, N. Szalai\altaffilmark{1},
 A. Szing\altaffilmark{1} } 
\email{kun@konkoly.hu}
\altaffiltext{1}{Konkoly Observatory, H-1121 Budapest, Konkoly Thege \'ut 15--17, Hungary}
\altaffiltext{2}{Instituto de Astrof\'{\i}sica de Canarias, E-38200, La Laguna, Tenerife, Canary Islands, Spain}
\altaffiltext{3}{Steward Observatory and Lunar and Planetary Laboratory, The University of Arizona, Tucson, AZ 85721, USA}
\altaffiltext{4}{Baja Astronomical Observatory of B\'acs-Kiskun County, PO Box 766, 6500 Baja, Hungary}

\date{Received / Accepted}

\begin{abstract}  

We conducted a long-term optical photometric and spectroscopic monitoring of the 
strongly variable, accreting young sun-like star \halp11, associated with the dark cloud 
Lynds~1340, that exhibited large amplitude (5--6 magnitudes in the \ic\ band) 
brightness variations on 2--3 years timescales, flat spectral energy distribution (SED), 
and extremely strong ($300 \lesssim$ EW/\AA $\lesssim 900$) \ha\ emission. 
In this Letter we describe the basic properties of the star, derived from our 
observations between 1999 and 2011, and put into context the observed phenomena. 
The observed variations in the emission spectra, near-infrared colors, and SED suggest 
that \halp11 (spectral type: K7--M0) is an eruptive young star, possibly similar in 
nature to V1647~Ori: its large-scale photometric variations are governed by variable 
accretion rate, associated with variations in the inner disk structure. The star recently 
has undergone strong and rapid brightness variations, thus its further
observations may offer a rare opportunity for studying structural and chemical 
rearrangements of the inner disk, induced by variable central luminosity.
\end{abstract}

                                                                               
\section{Introduction}
\label{Sect_1}
The irregular, large-scale photometric and spectroscopic variations 
of young low-mass stars reflect violent physical processes, originating from interactions 
of the stars with their accretion disks, and leading to clearing of the disk within an
astrophysically short time-scale. Most of the disk material accretes onto the central star 
in episodic outbursts, characterized by strongly increased optical, infrared, and  X-ray 
brightness \citep[e.g.][]{Kastner,ARH10,Audard}, powered by the enhanced accretion rate. 
The varying accretion luminosity may induce changes in the inner disk structure, 
which result in changing circumstellar extinction \citep*{ABR08,Brittain}.
The spectacular brightenings of eruptive young stars offer a rare 
opportunity to understand the effects of outbursts on disk structure and evolution. 
The target of the present paper is a young star whose photometric variations are comparable
in amplitude with those of eruptive stars, but differ from them in timescale. 
[KOS94]~HA11 (Simbad designation)
was identified as a candidate young stellar object by \citet{KOS94}, during an
objective prism H$\alpha$ survey for young stellar objects in the molecular cloud 
Lynds~1340, located at a distance of 600~pc from the Sun. It is associated with 
the protostellar-like IRAS source 02283+7230, and appears as a 16th magnitude star 
in both the DSS1 and DSS2 red images, obtained on 1954 September 28 and 1993 August 22, 
respectively, at the equatorial coordinates RA(J2000)=$2^\mathrm{h}33^\mathrm{m}01\fs5$, 
D(J2000)=$+72\degr43\arcmin27\arcsec$. 
We noticed its peculiar behavior in 1999, when the star was too faint to be detected in 
an unfiltered image obtained with a 2.2-m telescope (Fig.~\ref{Fig_chart}). 
We started monitoring the star in 1999, with the aim of observing its expected 
reappearance, and understanding the physical processes behind the observed phenomena. 
In this Letter we describe the basic properties of the star,
determined from our observations. 

\section{Observations}
\label{Sect_2}

\begin{figure}[!ht]
\resizebox{\hsize}{!}{\includegraphics{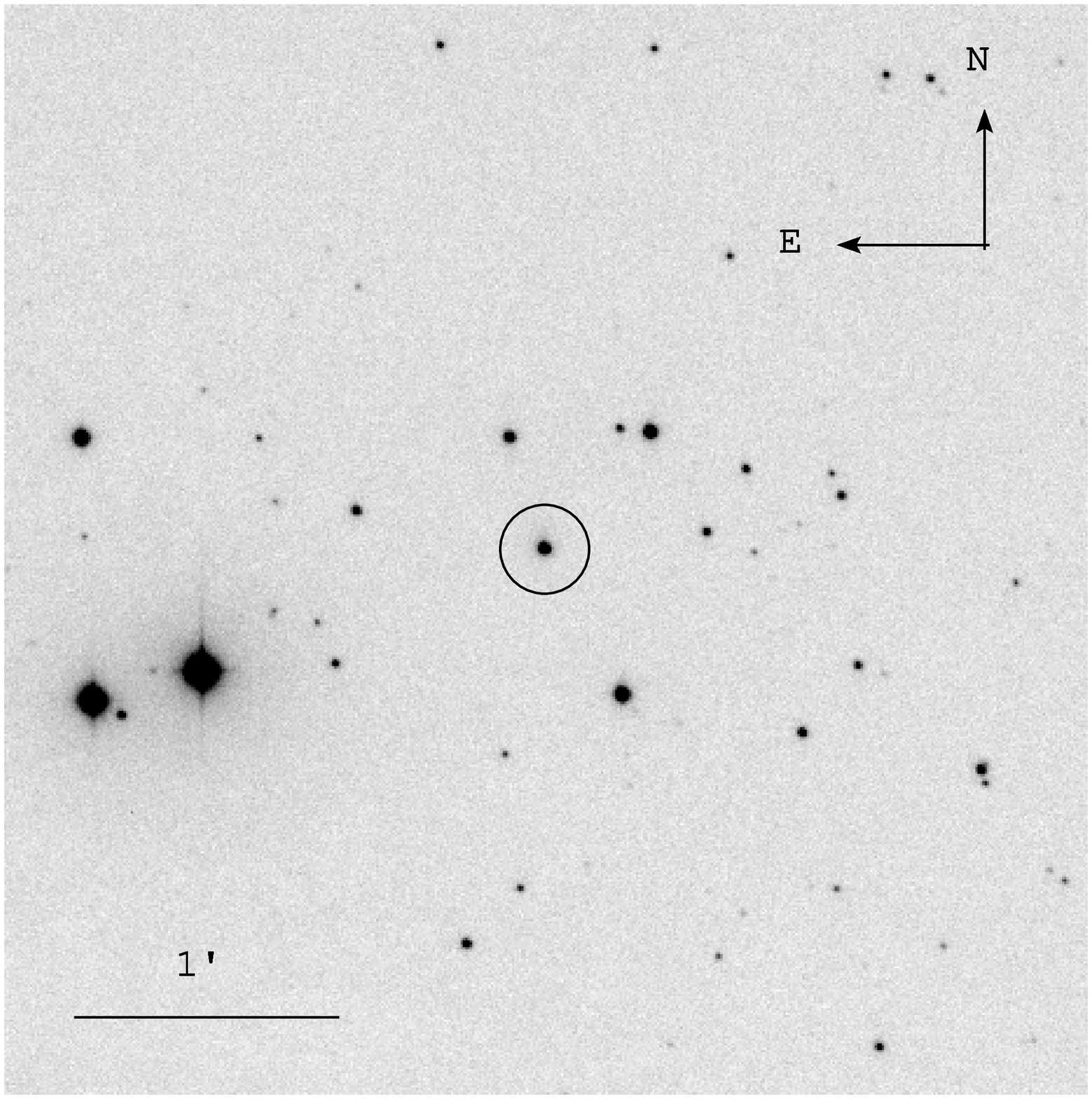}\includegraphics{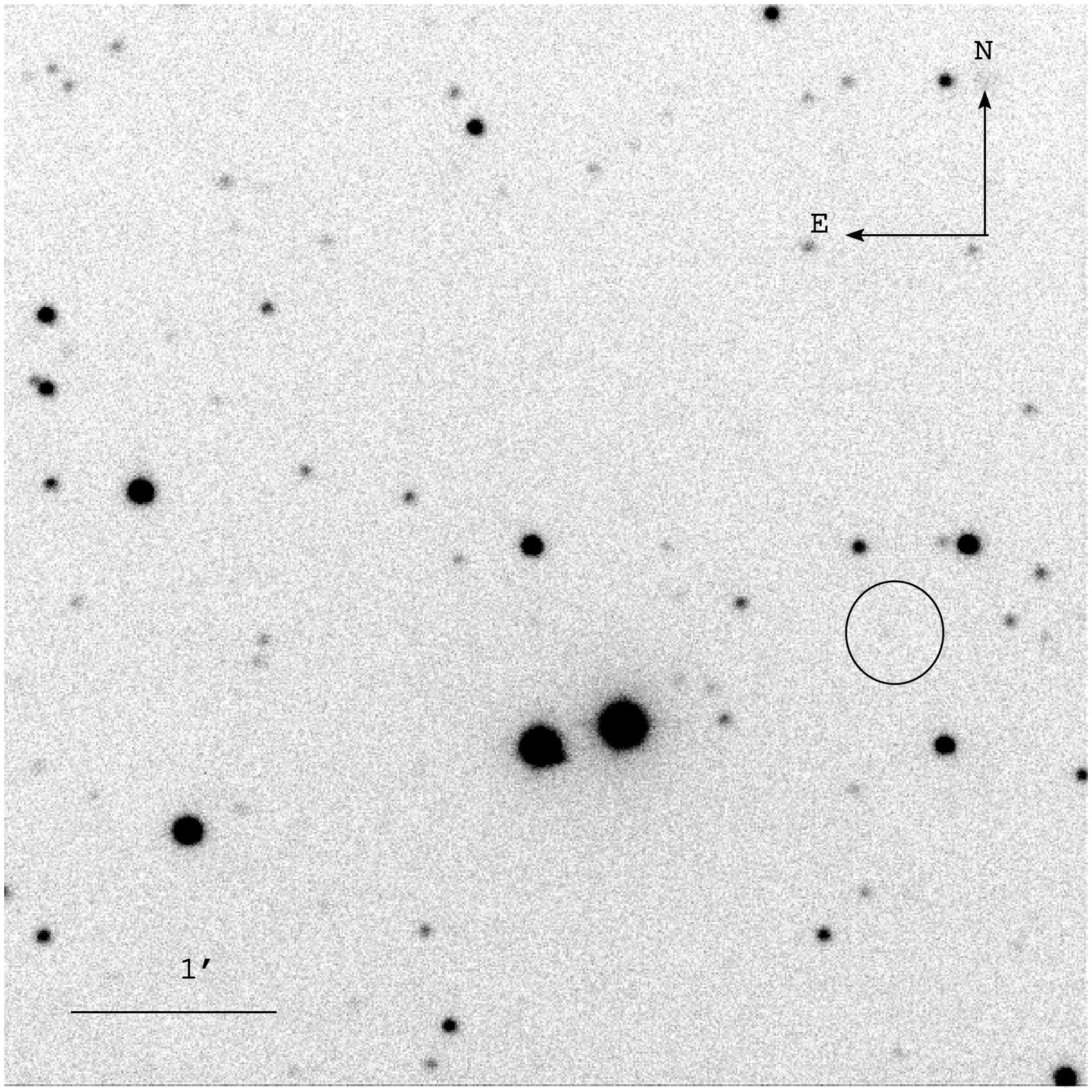}}
\resizebox{\hsize}{!}{\includegraphics{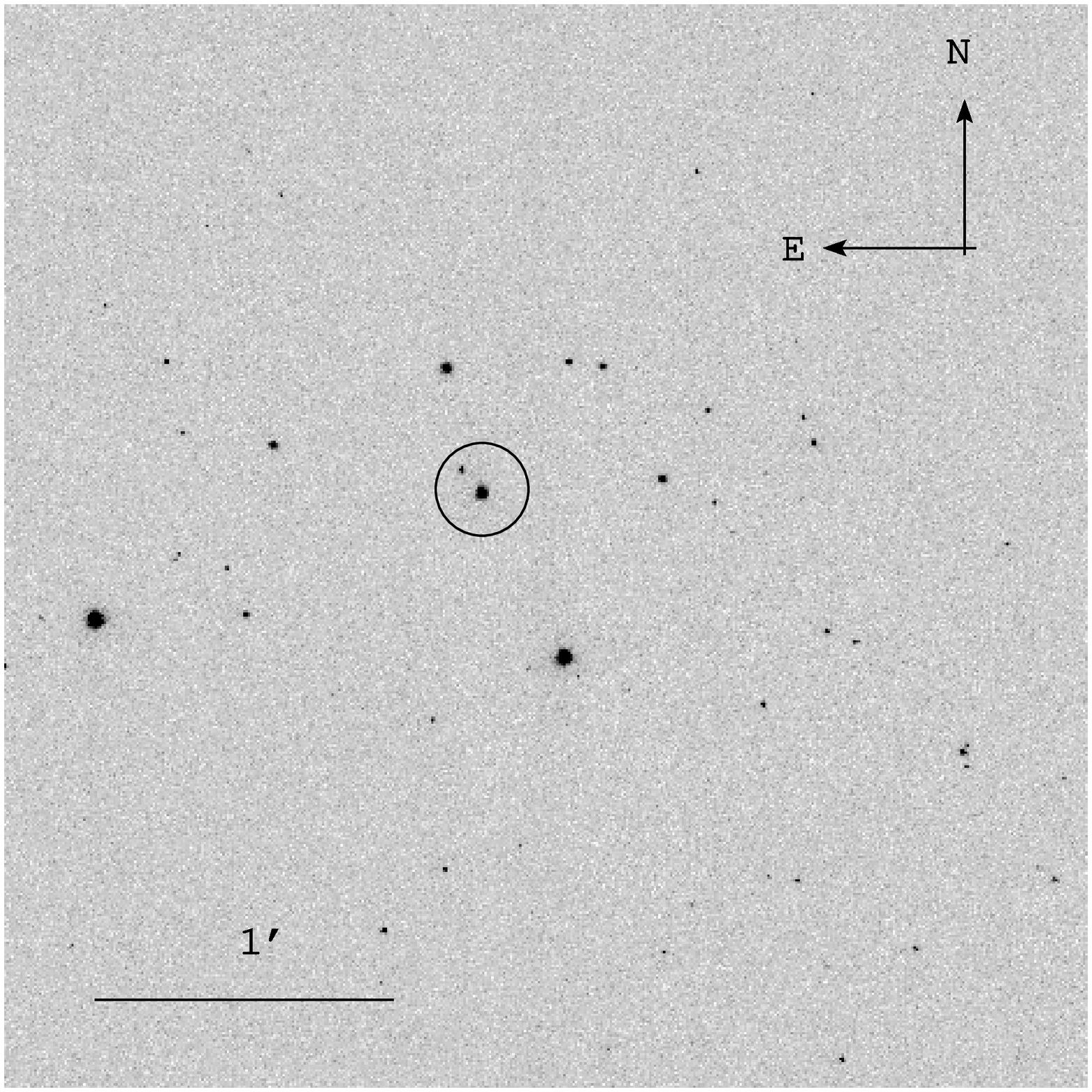}\includegraphics{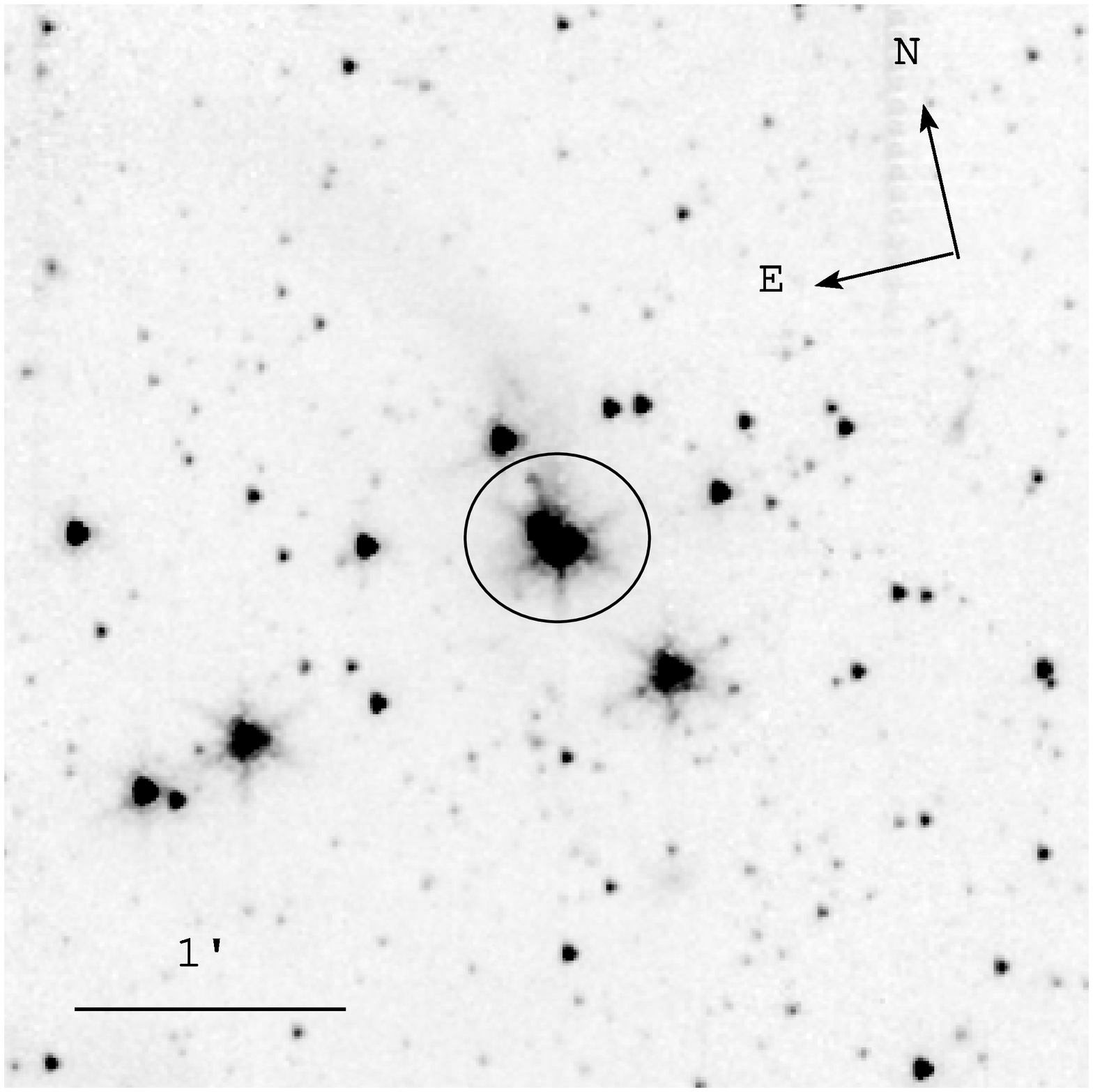}}
\caption{{\it Upper left}: \ic-band image of \halp11, obtained with the 2.2-m telescope 
of Calar Alto Observatory on 2009 October 9. Our target is the star marked by 
the circle.{\it Upper right}: An unfiltered image obtained with the same telescope 
on 1999 August 8. {\it Lower left}: \ks\ image obtained with NOTCam on 2007 October 1, 
showing also \halp11~B. {\it Lower right}: 3.6\,\micron \ IRAC image,
obtained on 2009 March 16.}
\label{Fig_chart}
\end{figure}

\paragraph*{Optical photometry} Photometric observations in the $VR_\mathrm{C}I_\mathrm{C}$ 
bands spanning the time interval between 1999 October and 2011 February  were 
performed with several instruments and telescopes. Most of the data were obtained 
with the 60/90/180~cm Schmidt telescope of the Konkoly Observatory, equipped with 
a Photometrics AT~200 camera before 2010 August and with an Apogee Alta U16 camera after
that date, and with the 1-m RCC telescope of the Konkoly Observatory, 
equipped with a Princeton Instruments VersArray:1300B camera. Further photometric 
observations were performed with the Electro Multiplying (EM) Andor Technology 
iXon$^\mathrm{EM}+888$  camera installed on the 50~cm Cassegrain telescope 
of the Konkoly Observatory, the `CAMELOT' CCD camera installed on the IAC--80 telescope 
of the Teide Observatory (Spain), and with the CAFOS instrument installed on the 
2.2-m telescope of the Calar Alto Observatory (Spain). More detailed description 
of these instruments is given in \citet{AP07}, \citet{Kospal11} and \citet{Kun11}. 
Typically three frames were taken with exposure times of 180--300\,s/frame each
night. The instrumental {\it V(RI)}$_{\mathrm{C}}$ magnitudes of \halp11 were determined by
aperture photometry using the `daophot' package in {\sc IRAF}. To transform the instrumental 
magnitudes into the standard system, we calibrated 
8 stars in the field of view of the 1-m telescope ($7\arcmin\times7\arcmin$). 
Calibration was made during five photometric 
nights. Standard stars in NGC~7790, published by \citet{Stetson}, were used as 
reference. On 2010 December 31 and 2011 January 26 we performed photometric  
observations of \halp11 through $i^{\prime}$ and $r^{\prime}$ filters, using WFGS2, 
installed on the 2.2-m telescope of University of Hawaii. The instrumental magnitudes were 
transformed into the standard $r^{\prime}i^{\prime}$ system using the $i^{\prime}$ and 
$r^{\prime}$ magnitudes of eight nearby stars, found in the {\it SDSS}\footnote{Funding 
for SDSS-III has been provided by the Alfred P. Sloan Foundation, the Participating Institutions, 
the National Science Foundation, and the U.S. Department of Energy. The SDSS-III web 
site is http://www.sdss3.org/. SDSS-III is managed by the Astrophysical Research Consortium 
for the Participating Institutions of the SDSS-III Collaboration including the University 
of Arizona, the Brazilian Participation Group, Brookhaven National Laboratory, University 
of Cambridge, University of Florida, the French Participation Group, the German 
Participation Group, the Instituto de Astrofisica de Canarias, the Michigan State/Notre 
Dame/JINA Participation Group, Johns Hopkins University, Lawrence Berkeley National Laboratory, 
Max Planck Institute for Astrophysics, New Mexico State University, New York University, 
Ohio State University, Pennsylvania State University, University of Portsmouth, Princeton 
University, the Spanish Participation Group, University of Tokyo, University of Utah, Vanderbilt 
University, University of Virginia, University of Washington, and Yale University.}
archive, and then were transformed into $R_\mathrm{C}$ and $I_\mathrm{C}$ using the formulae 
of \citet{Ivezic}. \halp11 itself was observed by the {\it SDSS} on 2005 November 3. The $i^{\prime}$
data point, transformed into \ic, is also included in our light curve.
Finding charts for \halp11 are shown in Fig.~\ref{Fig_chart}. 
The results of the photometry are listed in Table~\ref{Tab1},  
and the light curve in the \ic\ band is plotted in Fig.~\ref{Fig_lc}. 

\paragraph*{Optical spectroscopy} Intermediate resolution optical spectra of [KOS94]~HA11 were 
obtained in 2003 February 5, 2005 September 12, 2008 August 28, 31, and 2009 October 9 and 13, 
using CAFOS on the 2.2-m telescope of the Calar Alto Observatory. 
Both the R--100 grism, covering the 5800--9000\,\AA\ wavelength range, and the G--100 grism, 
covering the 5800--9000\,\AA\ wavelength range were used in 2008 and 2009. We used 
only G--100 in 2003, and only R--100 in 2005. The spectral resolution, using a 1.3~arcsec
slit width, was $\lambda / \Delta \lambda\approx 3500$ at $\lambda=6600$\,\AA.   
Broadband $VRI$ photometric images were taken immediately before 
the spectroscopic exposures for flux calibration. \halp11 was observed on 2007 September 12
using the double-armed, medium-resolution spectrograph ISIS, installed on the 
William Herschel Telescope  at the Observatorio del Roque de Los Muchachos, Canary Islands, 
Spain, with the grating R600R (0.49\,\AA/pixel). We observed our target with the slitless 
spectrograph WFGS2, installed on the 2.2-m telescope of the University of Hawaii 
on 2010 December 31. We used the low-dispersion grism~1 (3.8\,\AA/pix), 
and a wide \ha\ filter to isolate a $\pm250$~\AA\ wide region around 
the \ha\ line. We reduced and analysed the spectra in {\sc IRAF}. The overall appearance of 
the optical spectrum can be seen in the left panels of Fig.~\ref{Fig_spec}, and some variable
features are shown in the right panels. 

\begin{figure*}
 \centerline{\includegraphics[width=8cm]{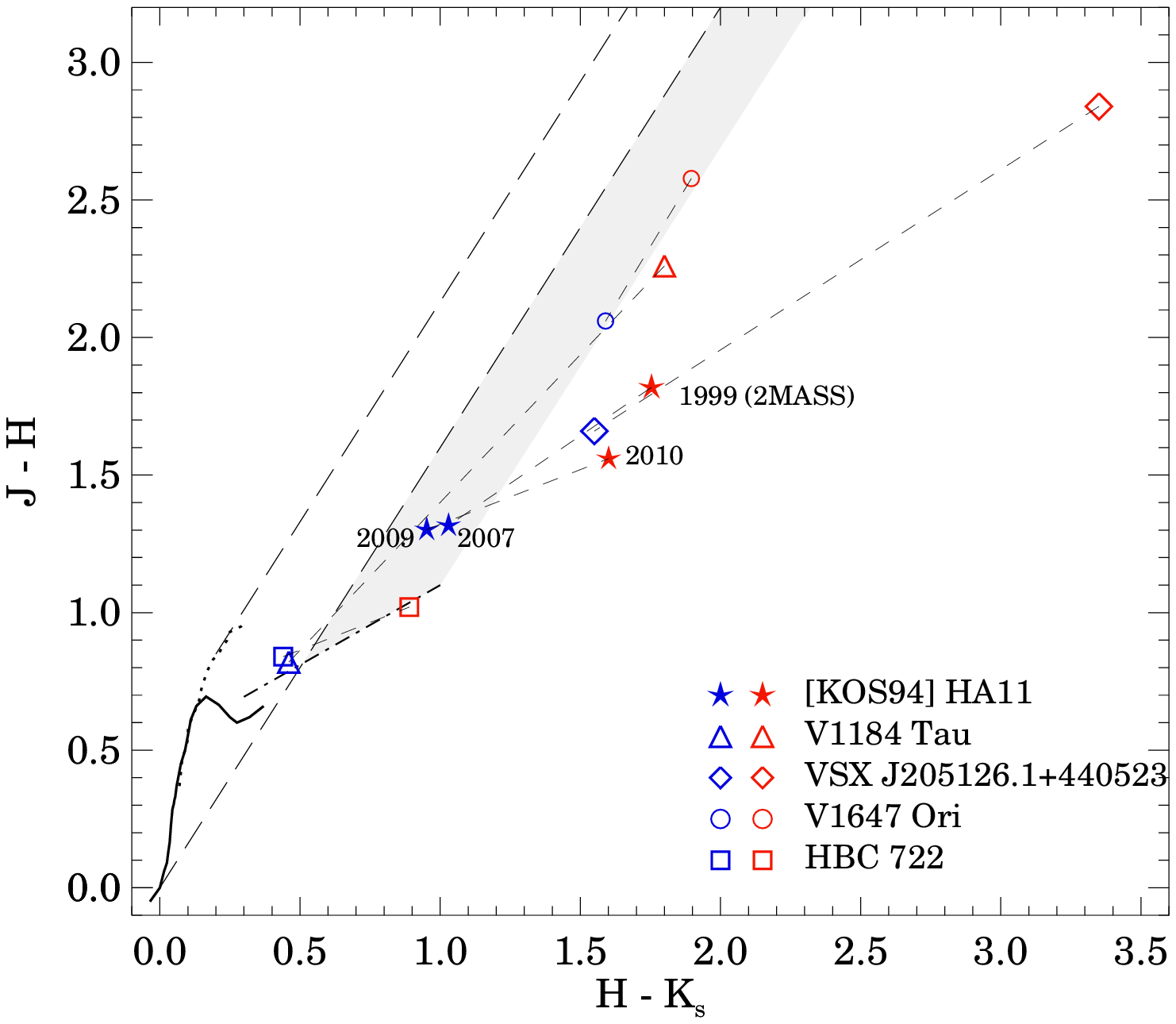}\includegraphics[width=8cm]{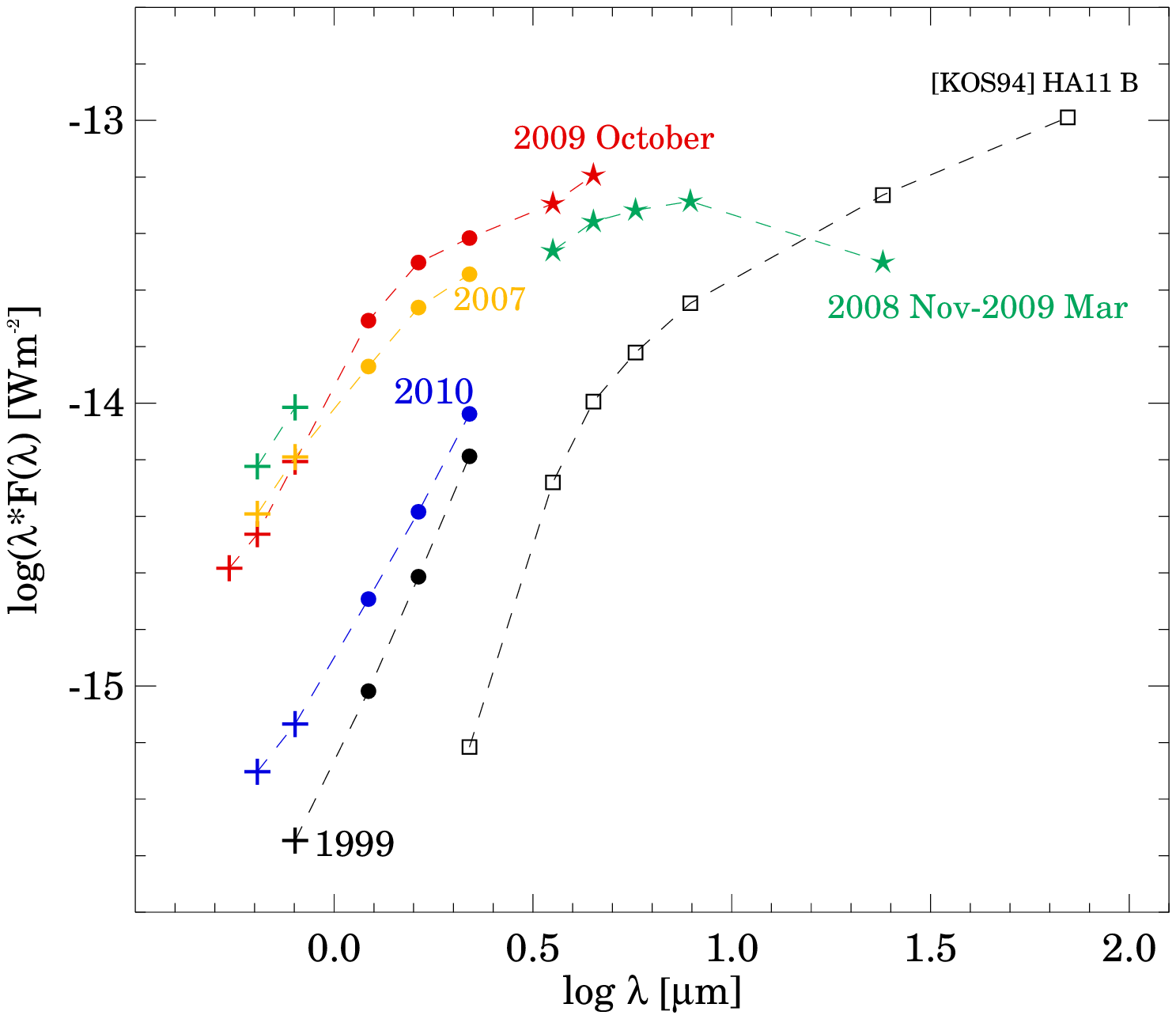}}
\caption{{\it Left}: Position of \halp11 $J-H$ vs. $H-K_\mathrm{s}$ diagram. The solid curve 
indicates the colors of the zero-age main sequence, and the long-dashed lines show the reddening
path \citep{Cardelli}. Dash-dotted line is the locus of unreddened T~Tauri stars \citep{Meyer97}, 
and the gray shaded band indicates the area occupied by reddened pre-main-sequence stars. Color
variations associated with the large-scale photometric variations of V1647~Ori \citep{AP07},  
V1184~Tau \citep{Grinin09}, HBC~722, and VSXJ205126.1+440523 \citep{Kospal11} are indicated
by blue (bright state) and red (faint state) symbols, connected by dashed lines.
{\it Right}: Spectral energy distribution of \halp11  for 1999, 2007, 2009, and 2010 October. Plus 
signs indicate our optical data, filled circles show the near-infrared observations, and
star symbols show the {\it Spitzer} measurements. Open squares show the SED of \halp11~B.}
\label{Fig_ir}
\end{figure*}

\paragraph*{Near-infrared photometry} $JHK_\mathrm{s}$ imaging observations of \halp11 
were performed on 2007 October 1, using the infrared camera NOTCam, installed on the
Nordic Optical Telescope\footnote{The Nordic Optical Telescope is operated on the island 
of La Palma jointly by Denmark, Finland, Iceland, Norway, and Sweden, in the Spanish 
Observatorio del Roque de los Muchachos of the Instituto de Astrof\'{\i}sica de Canarias.},
on 2009 October 12 using  NICFPS on the ARC~3.5-m telescope\footnote{The Apache Point 
Observatory 3.5-meter telescope is owned and operated by the Astrophysical Research Consortium.},
and on 2010 October 18 with MAGIC installed on the 2.2-m telescope 
of Calar Alto Observatory\footnote{The Centro Astron\'omico Hispano Alem\'an (CAHA) 
at Calar Alto is operated jointly by the Max-Planck-Institut f\"ur Astronomie and the Instituto de 
Astrof\'{\i}sica de Andaluc\'{\i}a (CSIC).}. The images were reduced in {\sc IRAF}. 
$JHK_\mathrm{s}$ magnitudes were derived by aperture photometry, and were transformed into 
the standard system by comparing them with the 2MASS magnitudes of four stars in the image 
field. The results of the $JHK_\mathrm{s}$ photometry are listed in Table~\ref{Tab1}, and plotted in
Fig.~\ref{Fig_ir}, together with the 2MASS data, obtained on 1999 December 12. 
In addition to the near-infrared counterpart of \halp11, 2MASS 02330153+7243269, the 
\ks\ images reveal an optically invisible companion at an angular distance of 6.2~arcsec 
NW from the star, 2MASS~02330247+7243315 (Fig.~\ref{Fig_chart}).

\paragraph{\textit{Spitzer} photometry} L\,1340 was observed by the \textit{Spitzer Space 
Telescope} using  IRAC on 2009 March 16 and MIPS on 2008 November 26 (Prog. ID: 50691, P.I. 
G. Fazio). Moreover, we conducted a monitoring of [KOS94]~HA11 with IRAC 
during the post-helium phase between 2009 October 24 and November 7 (Prog. ID: 60167, 
P.I. P. \'Abrah\'am). The data reduction procedures were same as described in detail 
in \citet{Kun11}. The results of the photometry of \halp11 are listed in Table~\ref{Tab1}. 
The fluxes of the companion are as 
follows: $F(3.6)=6.5\pm0.2$~mJy, $F(4.5)=15.7\pm0.5$~mJy, $F(5.8)=29.9\pm0.9$~mJy, 
$F(8.0)=61.0\pm1.9$~mJy,  $F(24)=435.9\pm17.4$~mJy, and $F(70)=2389.6\pm167.7$~mJy. 
The spectral index $\alpha=d(\log(\lambda F(\lambda)) / d\log\lambda = 1.88$  
between 2 and 24\,\micron\ reveals that it is a Class~I young stellar object. 
We assume that both stars are members of L\,1340. Their angular separation 
corresponds to a projected separation of 3600~AU. We refer to the companion as \halp11~B.
SEDs of both objects, based on our photometric results, are
plotted in the right panel of Fig.~\ref{Fig_ir}. 

\section{Results}
\label{Sect_3}

\subsection{Photometric behavior}
 
\begin{figure*}
\centerline{\includegraphics[width=16cm]{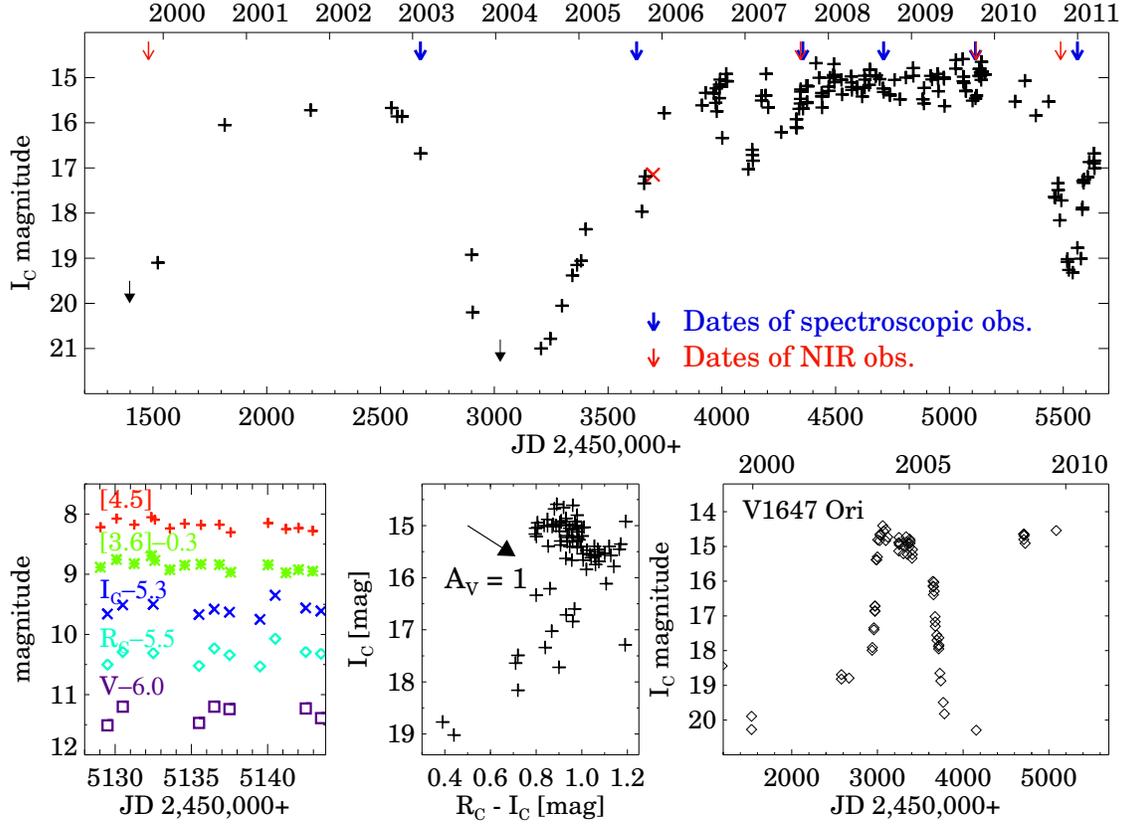}}
\caption{{\it Top}: \ic-band light curve of [KOS94]~HA11 for 1999--2011. The 
tips of the black downward arrows show the limiting magnitude of the image in which the 
target was not detected at the indicated date. Dates of the near-infrared photometric and optical 
spectroscopic observations are indicated. The red `x' indicates the {\it SDSS} measurement. 
{\it Bottom left}: Short-term multiwavelength variations observed between 2009 October 24 and 
November 7. {\it Bottom middle}: $I_\mathrm{C}$ magnitude of \halp11 as a function of the 
\rc$-$\ic\ color index. The arrow indicates the direction of the interstellar extinction.
{\it Bottom right}: \ic-band light curve of V1647 Ori for 2000--2010, based on data 
published by \citet{Briceno}, \citet{AP07}, \citet{ABR08}, \citet{Kun08}, \citet{AR09}, and 
\citet{Aspin09}.}
\label{Fig_lc}
\end{figure*}

\paragraph{Light curves} The upper panel of Fig.~\ref{Fig_lc} shows that the amplitude 
of the variations of \halp11 in the \ic\ band is about 6~mag, comparable with those of 
known eruptive young stars. The characteristic timescales of the faint 
and bright phases, however, differ from those of both FU~Ori (FUor) and EX~Lupi (EXor)
type eruptive stars. The star emerged from a deep brightness minimum in
2000, and stayed around \ic$\sim 16$~mag for two years. Its brightness dropped by a
factor of 100 in the \ic \ band in 2003, and then returned to the 2000--2002 level
by the end of 2005. In 2006--2007 the light curve was characterized by large-scale 
fluctuations. Between 2007 September and 2010 September the average brightness level was 
\ic=15.3~mag, and the average color index $R_\mathrm{C}-I_\mathrm{C}=0.96$~mag. 
The star faded to \ic=19.3 between 2010 October and mid-December, and then started brightening.
The short timescale behavior between 0.55 and 4.5 \micron\ is shown in 
the lower left panel of Fig.~\ref{Fig_lc}. Daily fluctuation on some 20\% flux 
level can be seen in each band. The fluxes at 3.6 and 4.5 \micron\ increased 
by some 50\% between 2009 March and October. These figures are consistent with the 
predictions of the model developed by \citet{Turner}, in which changing shadows, 
cast by dust clouds lifted by magnetic activity into the disk atmosphere are a cause of 
the daily to monthly mid-infrared variability. 
 
\paragraph{Color variations} The \ic\ vs. $R_\mathrm{C} - I_\mathrm{C}$
color--magnitude diagram (lower middle panel of Fig.~\ref{Fig_lc}) shows that during the 
brightness fluctuations in the bright state \halp11 tends to be redder when fainter: this 
behavior may result from either variable circumstellar extinction or variable hot 
spot coverage on the stellar surface due to fluctuating accretion rate \citep[e.g.][]{Scholz09}.
In dim state the star turned bluer, indicating that the fraction of scattered light in the optical
fluxes increased \citep[cf. e.g.][]{Bibo,Kun11}. The near-infrared color-color diagram 
(Fig.~\ref{Fig_ir}, left) shows that in the bright state (blue symbols) \halp11 appears a 
moderately reddened T~Tauri-type star, whereas in the low-state (red symbols) the near-infrared 
colors shift to the right, beyond the area occupied by reddened Class~II young stellar objects 
\citep*{Meyer97}. The color changes associated with the brightness variations 
differ from those originated from variable extinction, and are similar to those of the recently 
discovered eruptive young stars VSXJ\,205126.1+440523 and HBC\,722 
\citep{Kospal11}: the stars are shifted nearly parallel to the T~Tauri locus, 
indicating variations in the temperature and/or structure of the inner disks \citep{Meyer97}.

\paragraph*{The spectral energy distribution (SED)} of \halp11 for different epochs is 
shown in the right panel of Fig.~\ref{Fig_ir}. The slope of the bright-state SED between 
2 and 24\,\micron, $\alpha = -0.08$, classifies \halp11 as a {\it Flat SED} source \citep{Greene}. 

\subsection{Stellar properties}

\begin{figure}[ht]
\resizebox{\hsize}{!}{\includegraphics{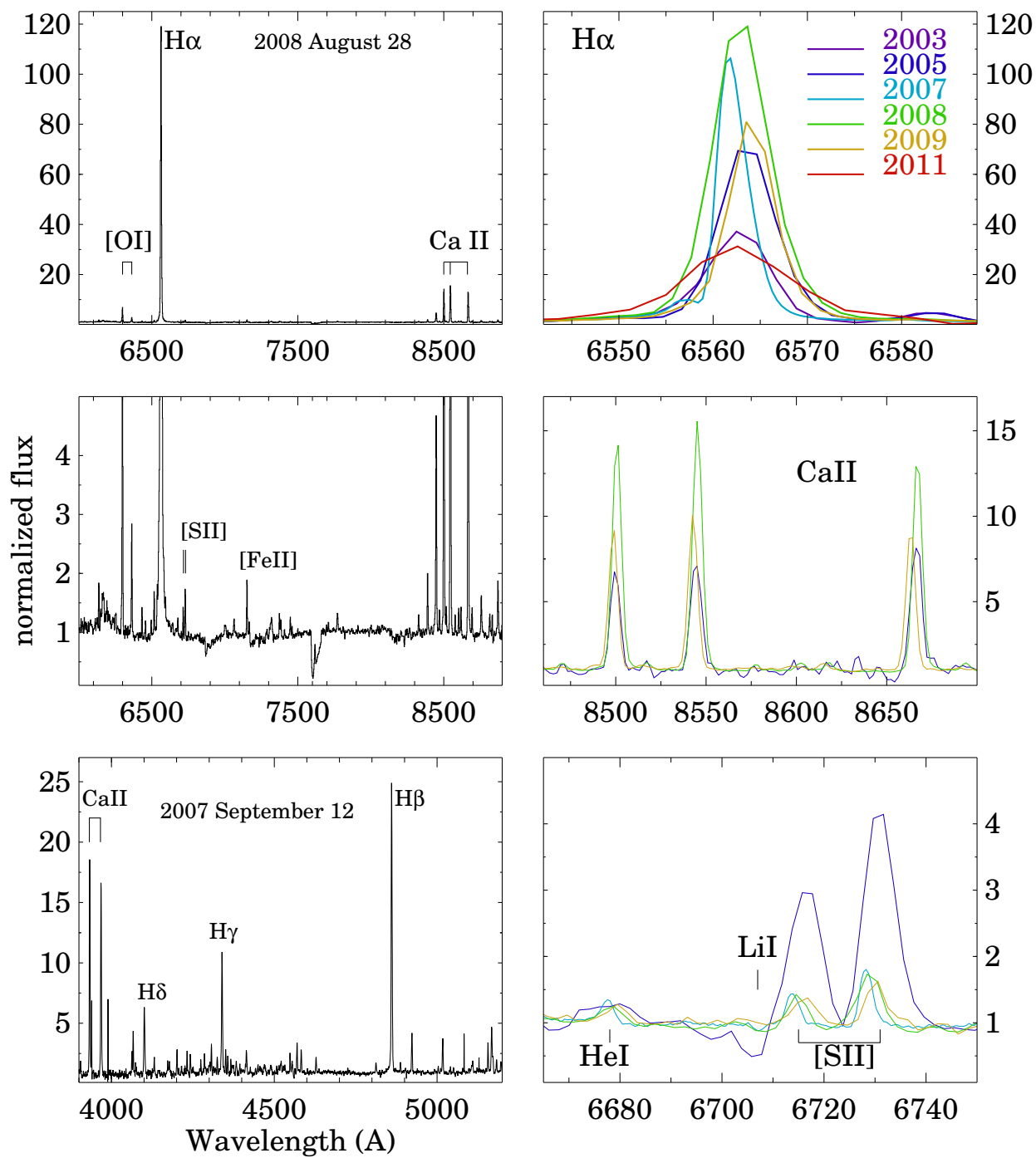}}
\caption{{\it Top and middle left\/}: Optical spectrum of \halp11 over the wavelength interval
6000--9000\,\AA, obtained on 2008 August 28, displayed on two intensity scales to show 
features of various flux level. {\it Bottom left\/}: The blue spectrum obtained on 2007 September 12. 
{\it Upper right\/}: Temporal behavior of the \ha\ line. {\it Middle right\/}: The \ion{Ca}{2} 
infrared triplet lines observed in 2005, 2008, and 2009. The color coding is same as in the 
upper right panel; {\it Bottom right\/}: Variability of the spectral 
region containing the \ion{He}{1}~6678, \ion{Li}{1}~6708, and [S~II]~6717,6731 lines.}  
\label{Fig_spec}
\end{figure}

\paragraph{Spectral classification} The left panels of Fig.~\ref{Fig_spec} suggest that 
\halp11 is a strongly accreting T~Tauri type star. The spectrum is dominated by the extremely 
strong  \ha\ line, and several permitted and forbidden emission lines, characteristic of 
strongly accreting T~Tauri stars, can be identified. The spectrum obtained on 
2005 September 12, when the star was some 3~mag fainter than in maximum, is less 
veiled than those obtained in the bright period between 2007 and 2009. 
The \ion{Li}{1} absorption line can be identified (Fig.~\ref{Fig_spec}, bottom right). 
We estimate a spectral type of K7--M0, based on the flux ratios $A$, $B$, and $C$, 
defined by \citet*{Kirk}, as well as on $R4/R7$ and $R6/R5$, introduced by \citet{Prosser}. 

\paragraph{Extinction and luminosity} We derive an extinction of $A_\mathrm{V}=4.3$\,mag 
towards the line of sight of \halp11 from the $I_\mathrm{C}-J = 2.09\pm0.04$ 
color index, averaged from the measurements in 1999, 2009, and 2010, assuming 
an unreddened $(I_\mathrm{C}-J)_0 = 0.77$ for the K7 spectral type \citep{KH95}. 
We determined the photospheric luminosity from the $J$ magnitude measured at the 
brightness minimum in 1999, after correcting for the above extinction, applying 
the bolometric correction $BC_J=1.54$, given by \citet*{Hartigan} for the K7 spectral type,
and a distance modulus of 8.89. The result is $L/L_{\sun}\approx0.07$, which, together 
with the effective temperature $T_\mathrm{eff}=4060$\,K of the K7 spectral type 
\citep{KH95}, places \halp11 near the main sequence in the HRD. The emission spectrum 
of the star contradicts this result, and suggests that the luminosity has been  
underestimated, most probably because the observed \ic\ and $J$ fluxes are dominated by 
scattered light, and thus the observed color indices underestimate the extinction. 
The implausibly low luminosity is indicative of high circumstellar extinction, originating
from an envelope or edge-on disk around the star \citep[e.g.][]{WH04,Huelamo}. 
 
\paragraph{Spectral variability} The temporal behavior of the \ha\ line is shown in the 
upper right panel of Fig.~\ref{Fig_spec}. The equivalent widths range from 300 
(2003 February 5) to 900\,\AA\ (2008 August 28). We measured EW(\ha)=530\,\AA \ at 
the latest brightness minimum on 2010 December 31. These values are unusually high for
classical T~Tauri stars \citep[cf.~e.g.][]{Fernan95, RPL96}, but occur in some
very low mass young stars \citep{Comeron}, and in Class~I stars \citep{WH04}. 
The shape of the \ha\ line is symmetric, except in the spectrum observed in 2007, when 
P~Cygni type absorption can be seen. The accretion-tracer \ion{Ca}{2} 
infrared triplet lines are also significantly weaker in the low-state spectrum in 
2005, than in the bright-state spectra (middle right panel of Fig.~\ref{Fig_spec}), 
suggesting that the photometric variations are associated with variable accretion 
activity \citep{Muzerolle98}. The \ion{He}{1} lines at 5873 and 6678\,\AA, apparent 
in the bright-state spectra, but absent from the spectrum obtained in 2005, 
lead to the same conclusion. Contrary to the permitted emission lines, the forbidden lines 
[\ion{O}{1}]~6300, 6363, [\ion{S}{2}]~6717,6731, and [\ion{Fe}{2}]~7154 exhibit
higher equivalent widths in the dim state, like in the spectra of V1647~Ori \citep{Fedele},
PV~Cep \citep{Kun11}, and V1184~Tau \citep{Semkov}. 

\paragraph{Accretion rates} The luminosity of the \ion{Ca}{2}\,$\lambda$\,8542 emission 
line is widely used for estimating the accretion rate, applyng the emprical relationship 
between the line luminosity and accretion luminosity  \citep[e.g.][]{Dahm08}. 
For lack of reliable extinction estimate, we can derive only a lower limit for 
the accretion rate, using $A_\mathrm{V}=4.3$ as lower limit for the extinction,
and assuming that the gas is accreted onto a 1~million year old K7~type star 
($M_{*}=0.8$\,M$_{\sun}$, $R_{*}=2$\,R$_{\sun}$). We obtain 
$\dot{M}_\mathrm{acc} > 1.6\times10^{-7}$\,$M_\sun/yr$ 
for both 2008 and 2009, and some ten times lower value for 2005.
These figures imply that the accretion luminosity in bright state is higher than 
the photospheric one, and changing accretion luminosity can account for most of 
the observed magnitude difference $\Delta I \sim 3$~mag between 2005 September and 2008 
October. The whole amplitude of the light curve would require some 100-fold 
variation in the accretion rate. 

\section{Possible nature of \halp11}
\label{Sect_4}

Large-amplitude optical photometric variations of young stars result from 
changing of either the accretion rate or the line-of-sight extinction. The near-infrared 
color-color diagram of \halp11 (Fig.~\ref{Fig_ir}) does not support the variable extinction scenario. 
The variable \ha, \ion{Ca}{2}, and \ion{He}{1} emission suggest that variable 
accretion may govern the photometric variability. Our results suggest that the accretion 
rate increased by an order of magnitude between 2005 September and 2008 August.
The P~Cygni absorption in the \ha\ line in 2007, which disappeared by 2008 August, 
indicated a temporary wind activity, probably associated with invigorated accretion. 
It resembles the behavior of the enigmatic eruptive star PV~Cep during 
the outburst in 1977--1979 \citep{Cohen81}. The bluer than average $I_\mathrm{C}-J$ 
color index, and the appearance of the \ion{He}{1} emission lines in the spectrum at 
the same epoch may suggest that a high-temperature component -- a hot region 
associated with an accretion shock -- appeared in the star-disk system in 2006--2007.

The brightening in 2000 might have resulted from a similar event, indicating 
repetitive outbursts. Exor outbursts are also repetitive, but
their light curves are characterized by several-year-long quiescent periods and 
weeks or months timescale outbursts. The bright and faint phases of \halp11 have 
comparable lengths. The morphology of the light curve resembles that of  
V1647~Ori (bottom left panel of Fig.~\ref{Fig_lc}), whose outbursts in 
2003--2005 and 2008 are separated by a three-year-long period of quiescence. 
Timescales of accretion bursts and the variations in the
star--disk system associated with the eruption depend on the physics of the
circumstellar environment of the stars. The increasing number of young eruptive stars 
apparently different from both FUors and EXors suggest that the present picture 
of the mechanisms of episodic accretion bursts still has to be refined. 
\halp11 may be a new member of yet unclassified eruptive young stars. The star 
currently undergoes significant brightness changes. Detailed monitoring across the 
recent period of high activity may give valuable information on the mechanism of outburst, 
and its effects on the disk structure and evolution.

\section*{Acknowledgements}

Our observations were supported by the OPTICON. OPTICON has received research 
funding from the European Community's Sixth Framework Programme under contract number
RII3-CT-001566. We thank Calar Alto Observatory for allocation of Director's 
discretionary time to this programme.  This work makes use of observations made with 
the \textit{Spitzer Space Telescope}, which is operated by the Jet Propulsion 
Laboratory, California Institute of Technology under a contract with NASA. 
Financial support from the Hungarian OTKA grant K81966 is acknowledged.

\clearpage

\clearpage

\clearpage

\clearpage

\clearpage

\begin{deluxetable}{lccccl}
\tabletypesize{\scriptsize}
\tablecolumns{6}
\tablewidth{0cm}
\tablecaption{Optical, near- and mid-infrared photometric observations of \halp11 between 
1999 and 2011 March.\tablenotemark{*} \label{Tab1}}
\tablehead{
\colhead{Date} & \colhead{MJD\tablenotemark{a}} &  \colhead{Band} & \colhead{Mag}
 & \colhead{$\Delta$ mag} & \colhead{Source\tablenotemark{b}} \\
\colhead{yyyymmdd}}
\startdata
19991028 &  1480.5 & J  	     & 16.49 & 0.13 & 2MASS \\  	    
19991028 &  1480.5 & H  	     & 14.68 & 0.06 & 2MASS \\  	    
19991028 &  1480.5 & K$_\mathrm{s}$  & 12.92 & 0.03 & 2MASS \\  	    
19991208 &  1521.5 & I$_\mathrm{C}$  & 18.55 & 0.05 & Schmidt/Photometrics \\  
20000927 &  1815.5 & I$_\mathrm{C}$  & 16.05 & 0.05 & Schmidt/Photometrics \\  
20011017 &  2193.5 & V  	     & 18.55 & 0.05 & RCC/Wright \\		    
20011017 &  2193.5 & R$_\mathrm{C}$  & 16.92 & 0.05 & Schmidt/Photometrics \\  
20011017 &  2193.5 & I$_\mathrm{C}$  & 15.72 & 0.05 & Schmidt/Photometrics \\  
\enddata  
\tablenotetext{a}{JD$-2,450,000$}
\tablenotetext{b}{Telescope/instrument or data base for the magnitude}
\tablenotetext{*}{The whole table is available only on-line as a machine-readable table.}
\end{deluxetable}

\end{document}